\documentclass[%
 aip,
 amsmath,amssymb,
 reprint,%
]{revtex4-1}

\usepackage{graphicx}
\usepackage{dcolumn}
\usepackage{bm}
\usepackage[utf8]{inputenc}
\usepackage[T1]{fontenc}
\usepackage{mathptmx}
\usepackage{etoolbox}
\usepackage[dvipsnames]{xcolor}

\usepackage{chemformula}

\definecolor{colorb}{rgb}{0, 0.4470, 0.7410}

\newcommand{\gp}{\dot\gamma}

\makeatletter
\def\@email#1#2{%
 \endgroup
 \patchcmd{\titleblock@produce}
  {\frontmatter@RRAPformat}
  {\frontmatter@RRAPformat{\produce@RRAP{*#1\href{mailto:#2}{#2}}}\frontmatter@RRAPformat}
  {}{}
}%
\makeatother
\begin{document}

\preprint{AIP/123-QED}

\title[Slow dynamics and time-composition superposition in gels of cellulose nanocrystals]{Slow dynamics and time-composition superposition in gels of cellulose nanocrystals}
\author{Lise Morlet-Decarnin}
 \email{lise.morlet-decarnin@ens-lyon.fr}
\author{Thibaut Divoux}%
\author{S\'ebastien Manneville}
\affiliation{Université de Lyon, Laboratoire de Physique, Ecole Normale Supérieure de Lyon, CNRS UMR 5672, 46 Allée d’Italie, 69364 Lyon cedex 07, France}

\date{\today}

\begin{abstract}
Cellulose nanocrystals (CNCs) are rodlike biosourced colloidal particles used as key building blocks in a growing number of materials with innovative mechanical or optical properties. While CNCs form stable suspensions at low volume fractions in pure water, they aggregate in the presence of salt and form colloidal gels with time-dependent properties. Here, we study the impact of salt concentration on the slow aging dynamics of CNC gels following the cessation of a high-shear flow that fully fluidizes the sample. We show that the higher the salt content, the faster the recovery of elasticity upon flow cessation. Most remarkably, the elastic modulus $G'$ obeys a time-composition superposition principle: the temporal evolution of $G'$ can be rescaled onto a universal sigmoidal master curve spanning 13 orders of magnitude in time for a wide range of salt concentrations. Such a rescaling is obtained through a time-shift factor that follows a steep power-law decay with increasing salt concentration, until it saturates at large salt content. These findings are robust to changes in the type of salt and in the CNC content. We further show that both linear and nonlinear rheological properties of CNC gels of various compositions, including, e.g., the frequency-dependence of viscoelastic spectra and the yield strain, can be rescaled based on the sample age along the general master curve. Our results provide strong evidence for universality in the aging dynamics of CNC gels, and call for microstructural investigations during recovery as well as theoretical modelling of time-composition superposition in rodlike colloids. 
\end{abstract}

\maketitle

\section{Introduction}

Colloidal gels raise great interest for their multiple applications in the design of soft materials \cite{Nelson.2019,Cao:2020,Diba:2017,Wang:2008}. Gels are formed through the percolation of attractive particles into a space-spanning network, which confers upon them an elastic response under small strains \cite{DelGado:2016,Rocklin:2021}. In practice, the manufacturing process of gel-based materials, for instance through 3D-printing \cite{Compton.2014,Siqueira.2017,Hausmann:2018}, generically involves shear flows that fully disrupt the gel microstructure, followed by a rest period during which the gel reforms. The structural build-up of the gel entails specific kinetics of the mechanical properties, eventually leading to the final product. From a more fundamental point of view, colloidal gels are intrinsically out-of-equilibrium systems whose physics still raise many open theoretical questions \cite{Bouzid:2020}. Therefore, it is essential to identify the key parameters that control the dynamics of colloidal systems following flow cessation. In most cases, such dynamics can be divided into two successive steps, respectively referred to as recovery and aging \cite{Joshi:2018}. First, starting from a fully fluidized suspension, the particles rapidly create new bonds right after flow cessation. Depending on the interparticle potential, colloids aggregate into either open, fractal-like cluster networks or thicker, glass-like bicontinuous structures, that constitute the backbone of the gel microstructure \cite{Weitz:1984,Zia:2014}. This sol-gel transition is characterized by relatively fast dynamics of the linear viscoelastic properties, where the elastic modulus $G'$ becomes larger that the loss modulus $G''$ typically over a few seconds to several minutes. Second, on longer time scales, the particles may rearrange locally and cooperatively due to thermal noise and to short-range, attractive interparticle forces, without any major large-scale change in the network structure \cite{Cipelletti:2000,Zia:2014,Bouzid:2017}. Such ``physical aging'' is associated with much slower dynamics of the viscoelastic properties that often take the form of a logarithmic time-dependence \cite{Derec.2003,Coussot:2006,Shahin:2012}, but may also follow other behaviors such as a power law \cite{Manley:2005,Negi:2009}.

Universality in the dynamical behavior of colloidal gels is probed by constructing master curves for the viscoelastic moduli $G'$ and $G''$, either during transient regimes including the above-described recovery and aging phases, or at steady state. These master curves are obtained through shifting a set of $(G',G'')$ curves generated by varying the system composition or some external control parameter, e.g., the temperature as commonly done for polymers \cite{VanGurp:1998,Larson:1999}. 
Master curves established at steady state, i.e., when aging can be neglected, focus on the viscoelastic spectra, $G'$ and $G''$ as a function of frequency $f$. Viscoelastic spectra are measured for various particle concentrations and/or interparticle potentials. They are subsequently collapsed onto a master curve by rescaling not only the frequency by a characteristic timescale, but also $G'$ and $G''$ both by the same characteristic modulus. Such master curves have been reported for a broad variety of colloidal suspensions, including fractal-like nanoparticles experiencing van der Waals attraction \cite{Trappe:2000}, spherical nanoparticles in depletion interaction \cite{Prasad:2003}, negatively charged rodlike nanoparticles with or without screened electrostatic interactions \cite{Lu:2014b,Huang:2021}, silica particles in various polymer solutions \cite{Pashkovski.2003,Adibnia:2017}, and more complex systems like caramel \cite{Weir:2016}, block
copolymer-cosolvent mixtures \cite{Krishnan:2010,Krishnan:2012}, and gluten protein gels \cite{Costanzo:2020}. These master curves hint at generic steady-state viscoelastic spectra that result from gel microstructures governed by similar topology and dynamics across a wide range of colloidal systems.

Searching for universality in the recovery and aging dynamics of colloidal gels raises more challenges as the viscoelastic spectra evolve over time \cite{Mours:1994,Geri:2018}. For time-dependent yet slowly evolving systems, the above rescaling can be adapted assuming quasi-stationarity to build a time-connectivity (or time-cure \cite{Weir:2016}) superposition principle, in which the characteristic timescale used to rescale the frequency is a function of the sample age,
as recently illustrated in aluminosilicate and silica gels \cite{Keshavarz.2021}.
Yet, a simpler way to deal with time dependence is to focus on a single frequency and to rescale the temporal evolution of the storage and/or loss moduli during the recovery and aging processes that follow flow cessation. Master curves obtained by shifting $G'(t)$ and/or $G''(t)$ in time only reflect the existence of a time-`\textit{parameter}' superposition principle, where the parameter can be the temperature, as reported in gels of silica particles \cite{Negi.2014}, the accumulated strain as shown in gels of cellulose nanocrystals mixed with an epoxide oligomer \cite{Rao.2019}, or the composition of the samples \cite{Cao.2010}. In the present work, we use this simple approach to unveil a time-composition superposition principle in colloidal gels made of anisotropic colloids, namely cellulose nanocrystals, in the presence of salt.

Cellulose nanocrystals (CNCs) are biosourced, biodegradable, and biocompatible nanoparticles. They have outstanding mechanical and optical properties, which make them relevant for the design of new green materials with numerous applications in medicine, electronics, food, and building industry \cite{Klemm.2018,Trache.2020,Lagerwall.2014,Li.2021}. CNCs come from the crystalline part of cellulose fibrils, which are extracted from diverse organic resources such as wood, cotton, algae, or some bacteria and mushrooms \cite{Klemm.2018}. They are rigid rodlike colloidal particles of length ranging between 100 and 500~nm and diameter between 5 and 20~nm depending on the source \cite{Lahiji.2010,Li.2021}. These particles are negatively charged due to the presence of sulfate groups on their surface. Therefore, when dispersed in water at weight fractions typically below 6~wt.~\%, they form stable suspensions thanks to repulsive electrostatic interactions. Adding salt in such aqueous suspensions induces screening of the electrostatic repulsion, and CNCs may subsequently aggregate to form various phases depending on the CNC and salt concentrations \cite{Xu.2020}. At low salt concentration, the phase diagram successively includes an isotropic liquid, a liquid crystalline phase, and a repulsive glass when the CNC concentration increases. At larger salt concentrations, colloidal gels are reported \cite{Cherhal:2015,Peddireddy:2016,Moud.2020} and give way to attractive glasses upon increasing the CNC concentration \cite{Xu:2019}.

Here, we focus on the slow aging dynamics of CNC gels following the cessation of a strong shear that rejuvenates the microstructure. By varying the salt concentration, we establish a time-composition superposition principle through the existence of a robust master curve for the temporal evolution of the storage modulus. We observe that the nature of the cation influences the gelation dynamics in a way that is compatible with the Hofmeister series. We further show that time-composition superposition allows one to rescale a number of linear and nonlinear rheological properties, based on the salt concentration, such as the frequency-dependence of the viscoelastic spectra, the loss factor, the characteristic strain at onset of nonlinear viscoelastic response as well as the yield strain. Our results therefore provide strong evidence for universality in the recovery and aging dynamics of CNC gels. 

\section{Materials and methods}

\subsection{Sample preparation}

Gels are prepared from a commercial CNC aqueous suspension provided by CelluForce (Montr\'eal, Canada) and containing 6.4~wt.~\% of CNCs extracted from wood (typical length 120~nm, and diameter 10~nm). The suspension is diluted to make samples containing 3.2 and 4.8~wt.~\% of CNCs. Gelation is induced by adding salt, either NaCl, KCl, or \ch{MgCl2} (Merck) at a concentration ranging from 5~mM to 240~mM.

The preparation protocol is as follows. ($i$)~We first homogenize the CelluForce aqueous suspension under high shear using mechanical stirring at 2070~rpm during 5~minutes (Turrax blender IKA RW20 equipped with an R1303 dissolver stirrer). ($ii$)~Salt is dissolved in distilled water at the desired concentration and the resulting solution is added to the CNC suspension under shear. ($iii$)~Mixing at 2070~rpm is then continued for 5 more minutes. ($iv$)~Finally, the sample is left at rest in a fridge for at least 24~hours. This protocol avoids the formation of large CNC agglomerates when adding salt and allows us to obtain homogeneous samples even at high salt concentrations. 

\subsection{Rheological measurements}
\label{sec:protocol}

The mechanical properties of the sample are measured using a stress-controlled rheometer (MCR 301 Anton Paar) equipped with a cone-and-plate geometry (rough cone of 40~mm diameter and 0.176~mm truncation). The smooth bottom plate is connected to a Peltier module, which sets the sample temperature to $T=23^{\circ}$C. In order to avoid evaporation, we use a home-made solvent trap made of a Plexiglas cylindrical dome covering the geometry, and we saturate the atmosphere surrounding the sample with water.

The following four-step rheological protocol is applied to all samples. ($i$)~The gel is presheared at $\dot{\gamma}_p=500$s$^{-1}$ during $t_p=20$~s to minimize the influence of previous mechanical history including the loading of the sample into the shear cell. ($ii$)~Preshear is stopped abruptly by setting the shear rate to zero, which defines the time origin $t=0$, and we subsequently measure the linear viscoelastic moduli, i.e., the storage modulus $G'$ and the loss modulus $G''$, every second for 1200~s by imposing small-amplitude oscillatory shear (SAOS) with strain amplitude $\gamma=0.2$~\% and frequency $f=$1~Hz.  ($iii$)~The viscoelastic spectrum [$G'(f), G''(f)$] is measured through SAOS with $\gamma=0.2$~\% by sweeping down the frequency $f$ logarithmically from 10~Hz to 0.1~Hz with 5 points per decade over a total duration of 270~s. ($iv$)~Finally, starting 1470~s after preshear cessation, we determine the yielding properties of the gel by sweeping up logarithmically the oscillatory strain amplitude from $\gamma=0.02$~\% to 500~\% at $f=1$~Hz, with 10 points per decade and a waiting time of 18~s per point, leading to a total duration of 790~s.

We checked that a strain amplitude of 0.2~\% lies within the linear viscoelastic regime for all samples so that the measurements of $[G'(t), G''(t)]$ in step ($ii)$ and of $[G'(f), G''(f)]$ in step ($iii$) do not interfere with the structural build-up and aging processes.

\section{Results}

\subsection{Gel recovery and aging following shear rejuvenation}

\subsubsection{Typical evolution of viscoelastic moduli and typical viscoelastic spectrum}
\label{sec:relax_gprime_gsecond}
\begin{figure}[b]
    \centering 
    \includegraphics[width=0.45\textwidth]{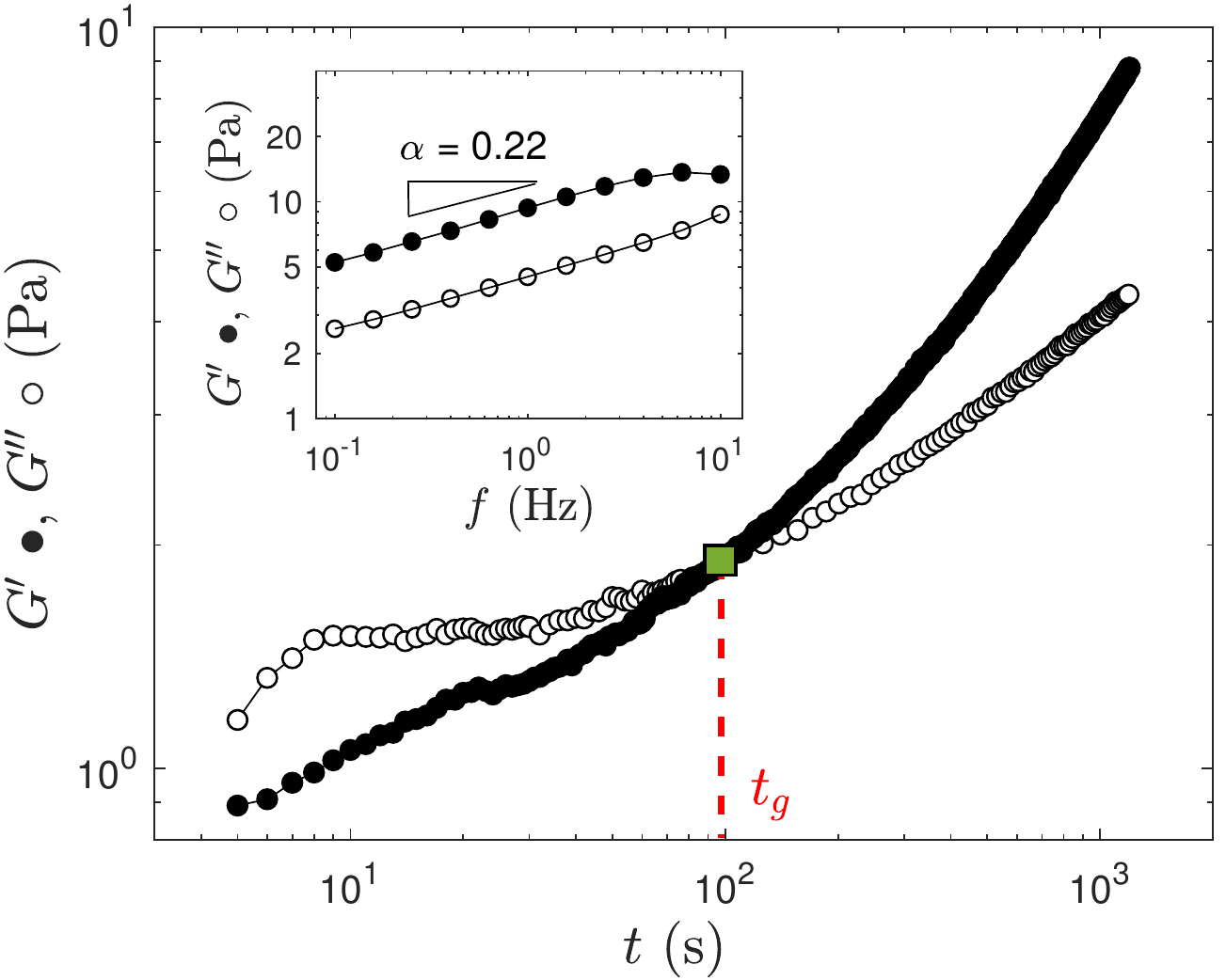}
\caption{Temporal evolution of the storage modulus $G'$ and the loss modulus $G''$ following a $20$~s preshear at $\gp_p=500$~s$^{-1}$. The red dashed line defines the time $t_g$ at which $G'$ and $G''$ cross each other, while the crossing point is marked by a green square. Experiment performed on a 3.2~wt.~\% CNC gel with 12~mM NaCl. Inset: Viscoelastic spectrum $G'$ and $G''$ as a function of frequency $f$ for a strain amplitude  $\gamma=0.2\%$, measured after the 1200~s rest period following preshear.}
    \label{fig:relax_gprime_gsecond_12mM_nacl}
\end{figure}

In order to study the recovery and aging kinetics of CNC gels, we focus on the evolution of the viscoelastic moduli $G'(t)$ and $G''(t)$ after shear rejuvenation, i.e., during step ($ii$) of the protocol described above. An example of this temporal evolution is shown in Fig.~\ref{fig:relax_gprime_gsecond_12mM_nacl} for a 3.2~wt.~\% CNC gel with 12~mM NaCl. Both the storage and loss moduli increase during the rest time after preshear. At short times, the loss modulus is larger than the storage modulus; hence, the sample behaves as a viscoelastic liquid up to $t=t_g\simeq 100$~s, where the two moduli take the same value. For $t>t_g$, the storage modulus $G'$ becomes larger than the loss modulus $G''$. Next, $G'$ keeps increasing faster than $G''$. Therefore, the sample behaves as a viscoelastic solid of ever-increasing elasticity, even beyond the 1200~s over which $G'(t)$ and $G''(t)$ are measured. Such an evolution from liquidlike to solidlike behavior is prototypical of the recovery and aging processes undergone by colloidal gels after flow cessation, as described above in the introduction. 

To further characterize the mechanical state of the sample, the inset of Fig.~\ref{fig:relax_gprime_gsecond_12mM_nacl} shows the viscoelastic spectrum $G'(f)$ and $G''(f)$ measured after the rest period of 1200~s, i.e., during step ($iii$) of the rheological protocol. Both the storage and loss moduli increase with frequency as weak power laws $G'\sim G''\sim f^\alpha$ of exponent $\alpha=0.22 \pm 0.02$. This means that the loss tangent $\tan \delta=G''/G'$ remains roughly constant and independent of $f$, which is reminiscent of a ``critical gel'' behavior \cite{Chambon:1987,Martin:1988,Winter:1997,Negi.2014}. Note, however, that due to the strong impact of the salt concentration on the kinetics reported below, some samples may undergo significant aging over the 270~s duration of the frequency sweep, so that one should remain cautious when interpreting these viscoelastic spectra. We also emphasize the fact that $t_g$ does not \emph{a priori} correspond to the gel point, which is actually defined as the point at which $\tan \delta$ first becomes independent of the frequency. A detailed, time-resolved study of the viscoelastic spectra using faster measurements over a wider range of frequencies thanks to an optimally windowed chirp sequence \cite{Geri:2018} is left for future work.

\subsubsection{Influence of the salt concentration}
\label{sec:serie_NaCl}

\begin{figure*}[!t]
    \centering 
    \includegraphics[width=1\textwidth]{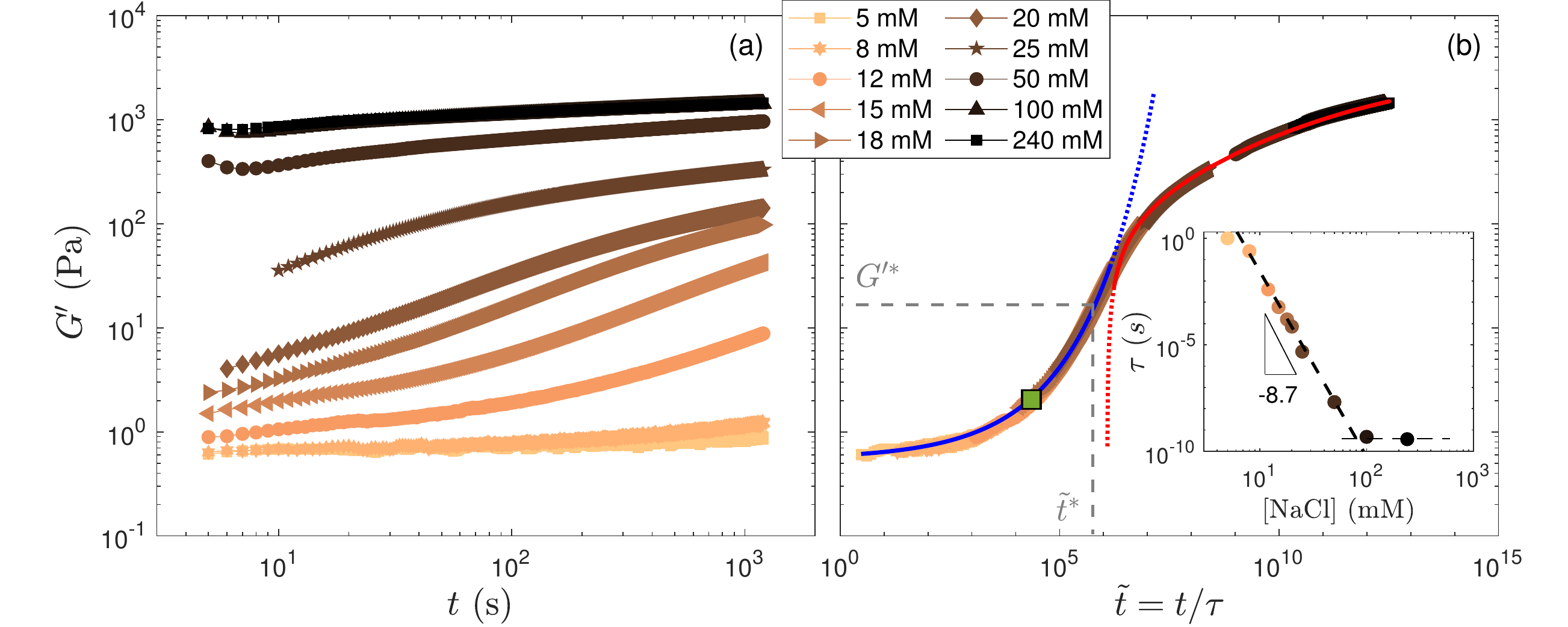}
    \caption{\label{fig:serie_NaCl} (a) Temporal evolution of the storage modulus $G'$ following a $20$~s preshear at $\gp=500$~s$^{-1}$. Experiments performed on 3.2~wt.~\% CNC gels with NaCl concentrations ranging from 5~mM to 240~mM (colored symbols). (b)~Master curve obtained by shifting the $G'(t)$ data in (a) along the time axis by a factor $1/\tau$. The response $G'(t)$ of the gel with 5~mM of NaCl is used as a reference ($\tau=1$~s). The green square shows the point at which $G'$ and $G''$ cross [see also Supplemental Fig.~S1(b)]. The gray dashed lines indicate the inflexion point of coordinates ($\tilde{t}^*$,$G'^*$). The blue and red lines show stretched exponential fits, respectively, $G'(\tilde{t} \lesssim \tilde{t}^*)=G'_0 \ \exp[(\tilde{t}/T_{p})^{p}]$ and $G'(\tilde{t} \gtrsim \tilde{t}^*)=G'_{\infty}(1-A \ \exp[-((\tilde{t}-\tilde{t}^*)/T_{q})^{q}])$ with $G'_0=0.55$~Pa, $T_{p}=8\times 10^3$, $p=0.28$, $G'_{\infty}=2200$~Pa, $A=1.11$, $T_{q}=8\times 10^{11}$, and $q=0.16$. Inset: Shift factor $\tau$ as a function of salt concentration.}
\end{figure*}

Figure~\ref{fig:serie_NaCl}(a) gathers the temporal evolutions of the storage modulus $G'(t)$ for 3.2~wt.~\% CNC gels with NaCl concentrations ranging from 5~mM to 240~mM and following the same preshear protocol as above. The corresponding loss moduli are shown in Supplemental Fig.~S1(a). In all cases, the storage modulus increases with time following preshear cessation. Yet, the kinetics strongly depend on the salt concentration: the higher the salt concentration, the higher the initial elasticity and the slower the subsequent growth of $G'(t)$. As discussed in the Supplemental Material, for NaCl concentrations above 18~mM, the gel is already in a solidlike state, i.e., $G'>G''$, only 5~s after preshear (first available data point for $G'$ and $G''$), so that the sol-gel transition cannot be resolved and only the slow aging is monitored. For salt concentrations larger than 100~mM, the evolution of the storage modulus appears to become independent of the NaCl content.

The shape of the viscoelastic responses for different salt concentrations prompts us to construct a master curve from the data in Fig.~\ref{fig:serie_NaCl}(a).  
By translating the $G'(t)$ curves in time by a factor $1/\tau$ that depends on the salt concentration, we obtain the master curve shown in Fig.~\ref{fig:serie_NaCl}(b). We arbitrarily take $\tau=1$~s for the lowest salt concentration of 5~mM, which thus constitutes a reference concentration. The corresponding data for the loss modulus $G''$ are presented in Supplemental Fig.~S1(b). Remarkably, the storage modulus increases over 13 decades of relative time $\tilde t=t/\tau$, and follows a sigmoidal curve in logarithmic scales with a clear inflexion point, which coordinates are denoted $\tilde{t}^*$ and $G'^*$ in the following. Over the 1200~s rest period investigated here, the inflexion point is only observed for salt concentrations in the range 15--20~mM [see middle curves in Fig.~\ref{fig:serie_NaCl}(a)].
Moreover, we note that the master curve cannot be fitted in logarithmic scales by a symmetric function with respect to the inflexion point. Rather, as shown by the red lines in Fig.~\ref{fig:serie_NaCl}(b), two separate stretched exponentials account respectively for the initial growth of the storage modulus ($\tilde{t}\lesssim \tilde{t}^*$) and for the late aging process ($\tilde{t}\gtrsim \tilde{t}^*$). 

The master curve obtained in Fig.~\ref{fig:serie_NaCl}(b) provides clear evidence for a time-composition superposition principle underlying the kinetics of the storage modulus: for any salt content, the time evolution of $G'(t)$ after shear rejuvenation can be mapped onto a segment of the master curve by simply rescaling the time by a factor that strongly decreases with the salt concentration. Thus, increasing the salt content corresponds to an increase of the effective age of the sample. In practice, our results imply that, provided one waits $3.2\times 10^{12}$~s, a 3.2~wt.~\% CNC gel containing 5~mM of NaCl should reach the same storage modulus as a 3.2~wt.~\% CNC gel with 240~mM of NaCl after 1200~s of rest. This suggests that the salt concentration only controls the kinetics of formation of a unique microstructure. 
Finally, the shift factor $\tau$, reported as an inset in Fig.~\ref{fig:serie_NaCl}(b), decreases as a power law of the salt concentration with exponent $-8.7\pm 0.2$, until a plateau is reached for salt concentrations above 100~mM. This plateau suggests that all the negative surface charges of the CNCs have been screened beyond 100~mM, and that any supplemental addition of salt poorly affects the viscoelastic properties of the CNC gel. However, the most salted samples still undergo significant aging since no saturation is reached in $G'(t)$ at the longest accessible times.

\subsubsection{Robustness of the master curve}
\label{sec:comparaison_sels}

\begin{figure*}[!t]
    \centering 
    \includegraphics[width=1\textwidth]{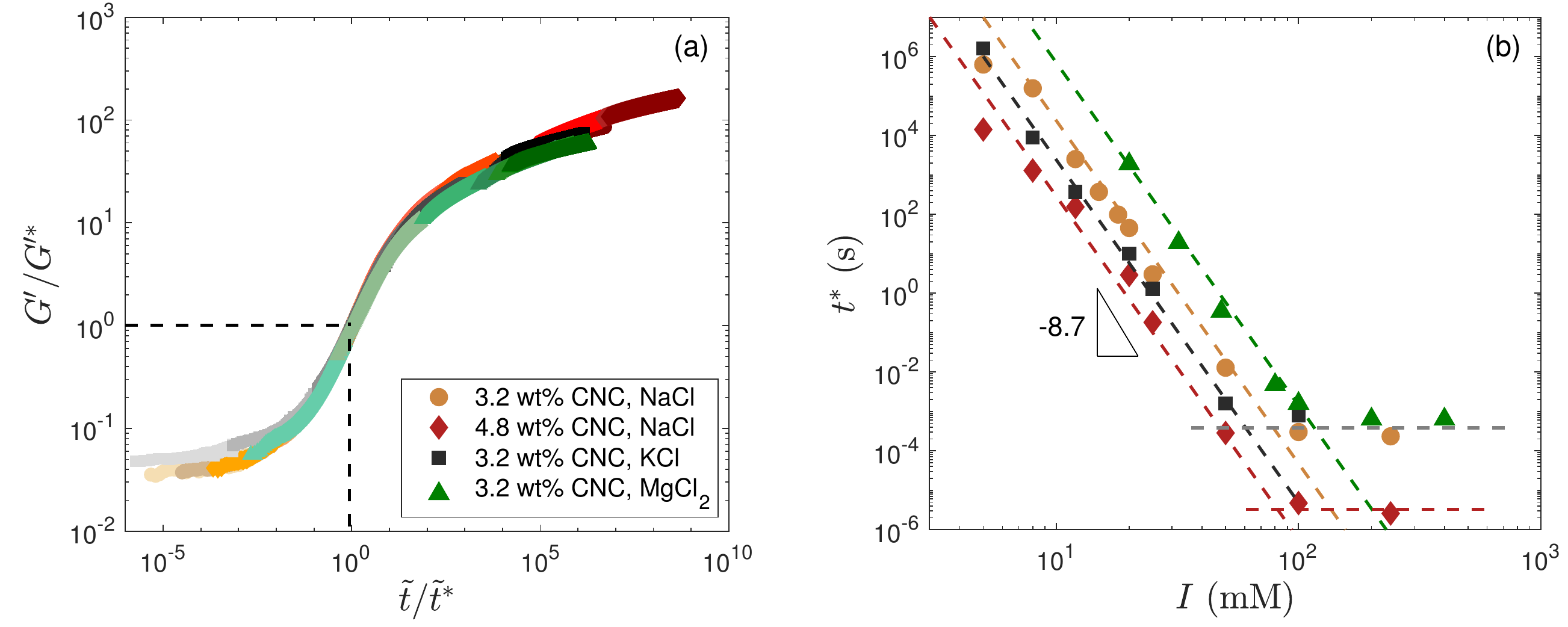}
    \caption{\label{fig:comparaison_sels} Time-composition superposition in CNC gels. (a)~Normalized elastic modulus $G'/G'^*$ vs.~normalized time $\tilde t/\tilde t^*$ during recovery and aging following a $20$~s preshear at $\gp=500$~s$^{-1}$. Experiments performed on four different series of samples containing either 3.2 or 4.8~wt.~\% CNC, and different types of salt, namely NaCl, KCl or \ch{MgCl2}, at concentrations ranging from 5~mM to 240~mM. The values of $G'^*$ and $\tilde t^*$ for each series of samples are reported in Table~\ref{tab:facteurs_renorm}. (b)~Inflexion time $t^*=\tau \times \tilde t^*$ vs.~the ionic strength $I$ [same symbols as in (a)]. Dashed lines correspond to power laws of exponent -8.7. Horizontal dashed lines highlight the plateaus reached beyond $I\simeq 100$~mM.}
\end{figure*}

In order to probe the robustness of the above-described rescaling, we now vary both the CNC concentration and the nature of the salt. In particular, the same preshear and rest protocol is used on 4.8~wt.~\% CNC gels with different concentrations of NaCl and on 3.2~wt.~\% CNC gels containing various concentrations of KCl or \ch{MgCl2}. For all samples, using the same procedure as in Sect.~\ref{sec:serie_NaCl}, the temporal evolution of the elastic modulus can be rescaled onto a master curve similar to the one presented above in Fig.~\ref{fig:serie_NaCl}(b), as shown in Supplemental Fig.~S2(a). Note that, for each series of gels, we take the gel with the lowest salt concentration as the reference for the time shift factor $\tau=1$~s [see Supplemental Fig.~S2(b)]. Further normalizing each $G'(\tilde t)$ data set based on the coordinates of the inflexion point ($\tilde t^*,G'^*$) leads to the general master curve reported in Fig.~\ref{fig:comparaison_sels}(a) (see also Supplemental Fig.~S3 for the $G''$ data). The four series of different CNC gels fall onto a unique master curve, which shape depends neither on the nature of the salt cation, nor on the CNC concentration. This demonstrates the robustness of time-composition superposition in the recovery and aging of CNC gels.

\begin{table}
\caption{\label{tab:facteurs_renorm} Coordinates ($G'^*$, $\tilde{t}^*$) of the inflexion point of each master curve [see Supplemental Fig.~S2 and normalized master curves in Fig.~\ref{fig:comparaison_sels}(a)] and parameters $B$ and $\beta$) of the best power-law fits $t^*=B/I^{\beta}$ of the inflexion time $t^*=\tau \times \tilde t^*$ vs. ionic strength $I$ for $I<100$~mM.}
\begin{ruledtabular}
\begin{tabular}{cccccc}
         CNC (wt.~\%) & Salt & $G'^*$ (Pa) & $\tilde{t}^*$ & B  & $\beta$\\
         \hline
         3.2 & NaCl & $17.0 \pm 0.5$ & $63 \pm 2 \times 10^4$ & 13.1 & $8.9$ \\
         4.8 & NaCl & $22.0 \pm 1.0$ & $14 \pm 1 \times 10^3$  & 11.2 & $8.5$ \\
         3.2 & KCl & $17.0 \pm 1.0$ & $16 \pm 1 \times 10^5$  & 12.1 & $8.7$\\
         3.2 & \ch{MgCl2} & $24.5 \pm 1.0$ & $19 \pm 2 \times 10^2$  & 14.6 & $8.8$\\
\end{tabular}
\end{ruledtabular}
\end{table}

The coordinates of the inflexion point used to normalize the various master curves are gathered in Table~\ref{tab:facteurs_renorm}. The values of $G'^*$ are all within the same order of magnitude and do not show any systematic variation with the nature of the cation or with the CNC concentration. The values of $\tilde{t}^*$, however, vary over three orders of magnitude, showing that both the nature of the cation and the CNC concentration strongly impact the time shift factor and thus the recovery and aging dynamics of CNC gels. In order to quantitatively compare the different series of gels, we consider the (dimensional) time $t^*=\tau \times \tilde t^*$ as a function of the ionic strength in Fig.~\ref{fig:comparaison_sels}(b). Here, the ionic strength $I$ is defined considering only the charges brought in solution by the introduction of salt, i.e., $I=C \times z^2$, where $C$ is the salt concentration, and $z$ is the salt valency. Multiplying $\tau$ by $\tilde t^*$ allows us to remove the possible influence of the different references used in the rescaling from one series to another. Actually, $t^*=\tau \times \tilde t^*$ corresponds to the time it would take for each gel to reach the inflexion point in its $G'(t)$ curve. Therefore, we shall refer to $t^*$ as the ``inflexion time'' in the following. Strikingly, Fig.~\ref{fig:comparaison_sels}(b) shows that the inflexion times all evolve in a similar manner with $I$: whatever the CNC concentration and the cation type, $t^*$ decreases as a power law of $I$ with an exponent of about $-8.7$, until it reaches a plateau beyond a critical ionic strength $I_c \simeq 100$~mM, when all negative CNC surface charges have been screened by the cations of the salt. For the sake of completeness, the parameters of the best power-law fits $t^*=A I^\beta$ below 100~mM are listed in Table~\ref{tab:facteurs_renorm}, although only power laws of exponent $-8.7$ are shown in Figs.~\ref{fig:serie_NaCl}(b) and \ref{fig:comparaison_sels}(b) for clarity. The mere difference between the various series is thus a vertical translation. We note that the kinetics become slower as one goes from \ch{K+} to \ch{Na+} and \ch{Mg^{2}+}, which follows the Hofmeister series (\ch{K+}>\ch{Na+}>\ch{Mg^{2}+})\cite{Kunz:2004a,Lyklema:2009}. Finally, increasing the CNC concentration from 3.2~wt.~\% to 4.8~wt.~\% accelerates the dynamics by almost two orders of magnitude.

\subsection{Scaling of the linear viscoelastic spectra based on time-composition superposition}
\label{sec:viscoelastic}

\begin{figure*}[t!]
    \centering 
    \includegraphics[width=1\textwidth]{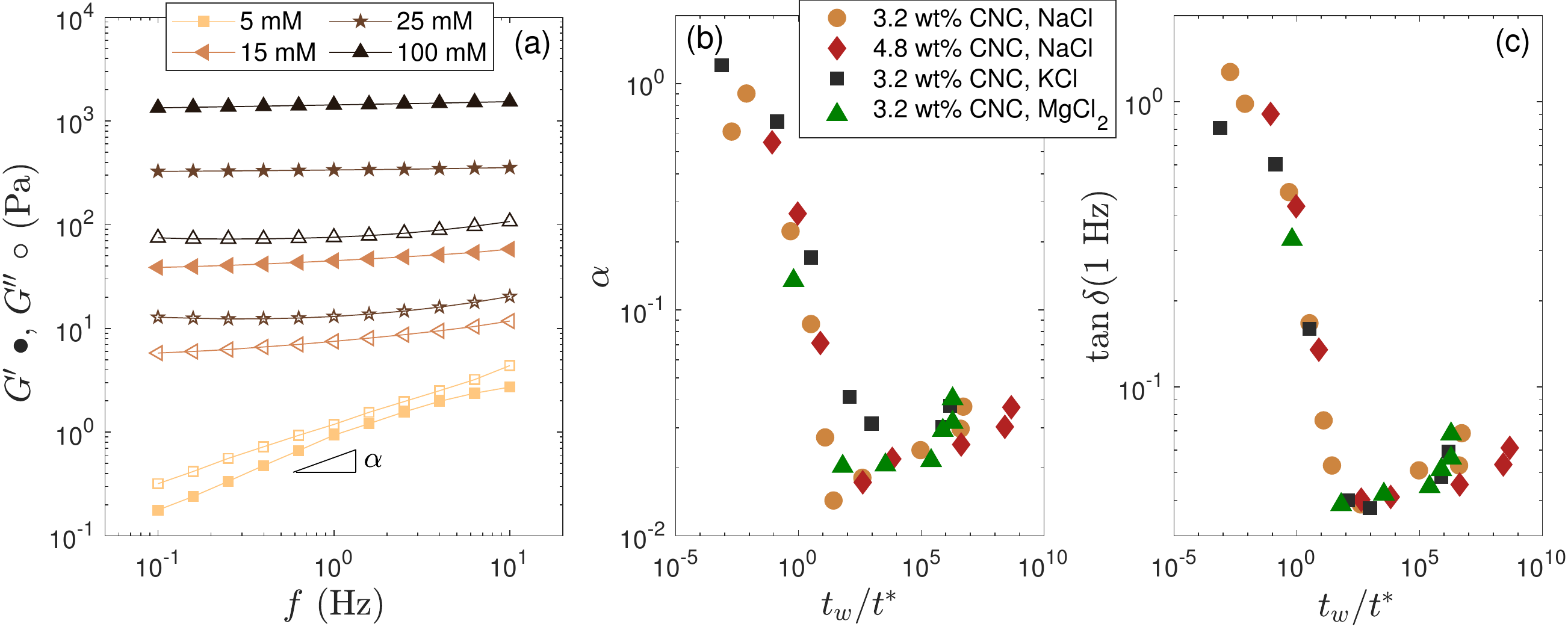}
    \caption{\label{fig:FS} (a)~Storage modulus $G'$ (filled symbols) and loss modulus $G''$ (open symbols) vs. frequency $f$ measured following $1200$~s of rest after a $20$~s preshear at $\gp=500$~s$^{-1}$ [step ($iii$) of the rheological protocol detailed in Sect.~\ref{sec:protocol}]. Experiments performed on 3.2~wt.~\% CNC gels with 5, 15, 25, and 100~mM NaCl. (b)~Exponent $\alpha$ extracted from power-law fits of the storage modulus, $G'(f)\sim f^\alpha$, over the whole frequency range, and (c)~loss tangent $\tan \delta$ at $f=1$~Hz vs. the effective age $t_w/t^*$ at the beginning of the frequency sweep test. Experiments performed on four different series of samples containing various CNC concentrations or types of salt as in Fig.~\ref{fig:comparaison_sels}.}
\end{figure*}

The time-composition superposition principle illustrated in Fig.~\ref{fig:comparaison_sels}(a) suggests that, within the range of samples investigated here, the gel properties should be the same for a given position along the master curve $G'(t)/G'^*$ vs. $\tilde t/\tilde t^*= t/t^*$, whatever the CNC concentration, the ionic force, and the type of salt. In order to test this hypothesis, we investigate the viscoelastic spectra [$G'(f), G''(f)$] measured after the 1200~s rest period following the cessation of preshear [step ($iii$) in the protocol of Sect.~\ref{sec:protocol}]. As mentioned above in Sect.~\ref{sec:relax_gprime_gsecond}, both $G'(f)$ and $G''(f)$ are well accounted for by similar power laws over the frequency range [0.1--10~Hz] (see inset in Fig.~\ref{fig:relax_gprime_gsecond_12mM_nacl} for the case of 12~mM NaCl). Figure~\ref{fig:FS}(a) shows additional viscoelastic spectra for NaCl concentrations ranging from 5~mM to 100~mM, and confirms that $G'(f)\sim f^\alpha$, yet with an exponent $\alpha$ that strongly depends on the salt content. We also note that, at large salt concentrations, the loss modulus $G''(f)$ does not follow such a clear power-law behavior as the storage modulus, and even seems to go through a minimum, a feature typical of soft glassy materials \cite{Mason:1995b,Mason:1995c,Purnomo:2006}. However, as already emphasized above, the time-dependence and aging of the gel may affect the measurements of the viscoelastic spectra, especially at low frequencies, which take longer to record. Therefore, here, we only focus on the exponent $\alpha$ inferred from $G'(f)$.

Figure~\ref{fig:FS}(b) and (c) respectively display the exponent $\alpha$ and the value of the loss tangent $\tan\delta=G''/G'$ at $f=1$~Hz measured for the various series of gels as a function of the normalized time $t_w/t^*$, where $t_w$ is the waiting time (or aging time) at rest after preshear cessation, here $t_w=1200$~s. Note that $t_w/t^*=\tilde{t}_w/\tilde{t}^*$ corresponds to the location of the start of the frequency sweep, namely to the end of the 1200~s rest period, along the master curve of Fig.~\ref{fig:comparaison_sels}(a) relative to the inflexion point. In other words, $t_w/t^*$ measures the ``effective age'' of the sample relative to the inflexion time. Remarkably, both observables $\alpha$ and $\tan\delta(1\,$Hz) follow a universal dependence on $t_w/t^*$, which provides very strong support for the time-composition superposition revealed in Sect.~\ref{sec:comparaison_sels}. More precisely, the exponent $\alpha$ strongly decreases from about 0.9 at the early stages of the effective dynamics, down to about 0.02 for $t_w\simeq 10\,t^*$ [see Fig.~\ref{fig:FS}(b)]. 

Concomitantly with the decrease in $\alpha$, the loss tangent at $f=1$~Hz drops with $t_w/t^*$ by more than one order of magnitude, from values slightly above 1 indicative of a viscoelastic liquid, down to about 0.04 signalling clear solidlike behavior [see Fig.~\ref{fig:FS}(c)]. Here again, the data for all series of samples nicely collapse onto a single curve, which confirms the time-composition superposition principle for linear viscoelastic properties. Finally, both observables saturate rather abruptly to a constant value at $t_w\gtrsim 10\,t^*$ and seem to increase for $t_w\gg t^*$, although the data are somewhat more scattered for large effective ages. Such a saturation does not imply that the sample microstructure no longer evolves, as $G'(t)$ is seen to keep increasing even at the longest times. It rather means that consolidation further takes place with a constant balance between elasticity and dissipation.

\subsection{Scaling of nonlinear viscoelastic parameters based on time-composition superposition}
\label{sec:nonlinear}

In order to go beyond linear viscoelasticity, we now turn to the gel response to large-amplitude oscillatory shear (LAOS) and ask whether the time-composition superposition principle also holds for nonlinear viscoelastic parameters. To address this question, after a rest period of 1470~s (that includes the previous frequency sweep test), the sample is submitted to an oscillatory strain at $f=1$~Hz, which amplitude $\gamma$ is increased from 0.02~\% to 500~\% [step ($iv$) in the protocol of Sect.~\ref{sec:protocol}].
\begin{figure*}
    \centering 
    \includegraphics[width=1\textwidth]{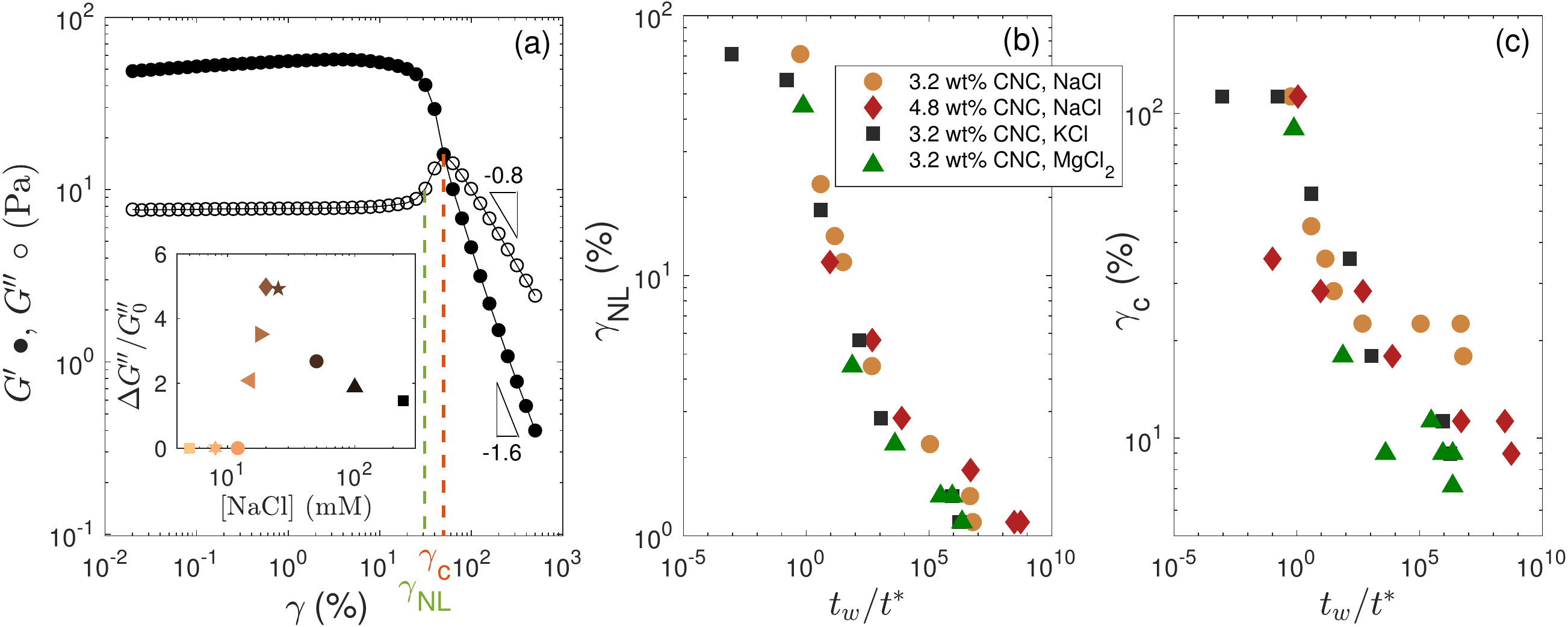}
    \caption{\label{fig:SS} (a)~Storage modulus $G'$ and loss modulus $G''$ vs. strain amplitude $\gamma$ following $1470$~s of rest after a $20$~s preshear at $\gp=500$~s$^{-1}$ [step ($iv$) of the rheological protocol detailed in Sect.~\ref{sec:protocol}]. The green dashed line defines the characteristic strain at onset of nonlinear response, corresponding to $G''(\gamma_{\rm NL})=1.1G''_0$, where $G''_0$ is the plateau value of the loss modulus at small strain amplitude. The red dashed line defines the yield strain $\gamma_{\rm c}$ at which $G'=G''$. Experiment performed on a 3.2~wt.~\% CNC gel with 15~mM NaCl. Inset: Dimensionless amplitude of the loss modulus overshoot $\Delta G''/G''_0$ vs. NaCl concentration. Experiments performed on 3.2~wt.~\% CNC gels with various NaCl concentrations. (b)~Characteristic strain $\gamma_{\rm NL}$ and (c)~yield strain $\gamma_{\rm c}$ vs. the effective age $t_w/t^*$ at the beginning of the LAOS test. Experiments performed on four different series of samples containing various CNC concentrations or types of salt as in Fig.~\ref{fig:comparaison_sels}.}
\end{figure*}
An example of the evolution of the storage modulus $G'$ and loss modulus $G''$ as a function of the strain amplitude $\gamma$ during the LAOS test is shown in Fig.~\ref{fig:SS}(a) for a 3.2~wt.~\% CNC gel with 15~mM NaCl. At small strains, in the linear regime, $G'>G''$ and the sample behaves as a soft solid (see also Supplemental Fig.~S4 for more examples). The moderate increase in $G'$ is due to the aging of the sample while the strain is being ramped up over 790~s. Such an increase is in agreement with the master curve of Fig.~\ref{fig:serie_NaCl}(b). When the strain reaches about 20~\%, the storage modulus drops abruptly, while the loss modulus goes through a maximum, hence following a so-called ``type III'' yielding scenario \cite{Hyun.2002,Hyun:2011}. This yielding response under LAOS is consistent with recent observations on CNC gels prepared in a similar concentration range \cite{Danesh:2021}. In practice, the characteristic strain $\gamma_{\rm NL}$ for which the loss modulus has increased by 10\% compared to its plateau value at low strain, is taken as the onset of nonlinear viscoelasticity.
Finally, for even higher strain amplitudes, the point at which the storage and loss moduli cross defines the yield strain $\gamma_{\rm c}$, beyond which the sample behaves as a viscoelastic liquid, with $G''>G'$. 

Figure~\ref{fig:SS}(b) and \ref{fig:SS}(c) respectively test the time-composition superposition on the two characteristic strains $\gamma_{\rm NL}$ and $\gamma_{\rm c}$ displayed as a function of the effective age $t_w/t^*$ at the beginning of the LAOS test, i.e., with $t_w=1470$~s. Whatever the nature of the cation or the CNC concentration, the strain amplitude $\gamma_{\rm NL}$ at the onset of the nonlinear regime shows the same decreasing trend when moving along the master curve of Fig.~\ref{fig:comparaison_sels} [see Fig.~\ref{fig:SS}(b)]. The yield strain $\gamma_{\rm c}$ follows a similar evolution, although the data are more dispersed  [see Fig.~\ref{fig:SS}(c)]. The latter observation suggests that both the nature of the counterion and the exact CNC content may have some non-negligible impact on the scaling of the yield point. Yet, the yielding transition around $\gamma_{\rm c}$ is a highly dynamical process and the fluidization scenario is likely to involve spatially heterogeneous dynamics, such as shear bands or fractures \cite{Perge:2014,Gibaud:2016,Gibaud:2020}. Therefore, a complex interplay between the sweep rate of the strain amplitude, the total duration of 790~s of the test, and the yielding dynamics under LAOS most probably accounts for the dispersion around a single curve in Fig.~\ref{fig:SS}(c). We conclude that the superposition principle also holds for nonlinear viscoelastic properties and that the present CNC gels become more and more sensitive to strain as their effective age increases.

Finally, we note that for $\gamma > \gamma_{\rm c}$, both $G'$ and $G''$ display a decrease that is well fitted by power laws $G' \sim \gamma^{-\nu'}$ and $G'' \sim \gamma^{-\nu''}$, with $\nu'=1.6\pm 0.2$ and $\nu''=0.8\pm 0.1$ in the case of Fig.~\ref{fig:SS}(a). Remarkably, both these exponents are insensitive to the type of counterion and to their concentration (see Supplemental Fig.~S5), while their ratio remains roughly constant to $\nu'/\nu'' \simeq 2$. This value is consistent with that reported for a broad range of soft glassy materials such as dense suspensions of hard spheres \cite{Miyazaki:2006} and soft particles \cite{Koumakis.2012,Migliozzi:2020}. In sharp contrast with $\nu'$ and $\nu''$, the amplitude of the peak in $G''$ at the yield point strongly depends on the salt content. The inset in Fig.~\ref{fig:SS}(a) shows the relative amplitude of the overshoot in the loss modulus, $\Delta G''/G''_0$, where $\Delta G''=G''(\gamma_{\rm c})-G''(\gamma_{\rm NL})$ and $G''_0$ is the plateau value at low strain amplitude, as a function of the NaCl concentration for 3.2~wt.~\% CNC gels. At low NaCl concentration, there is no overshoot in $G''$ and the sample is barely solidlike when the LAOS test is performed [see also Supplemental Fig.~S4(a)]. The maximum only appears beyond a salt content of 12~mM. Upon increasing the salt concentration, $\Delta G''/G''_0$ reaches a maximum for a concentration of about 20~mM [see also Supplemental Fig.~S4(b)], before decreasing and leveling off to a constant value of about 1.4 [see also Supplemental Fig.~S4(c)]. When plotted as a function of $t_w/t^*$ in Supplemental Fig.~S6, the overshoot in $G''$ is shown to follow a universal, non-monotonic behavior for all samples investigated here. Therefore, the nonlinear observable $\Delta G''/G''_0$ is also consistent with time-composition superposition.

\section{Discussion and Conclusion}

\subsection{Summary of the main results}

We have studied the recovery and slow aging dynamics of CNC gels following the cessation of a strong shear that rejuvenates their microstructure. We have shown that the dynamics strongly depend on the salt concentration: the storage modulus evolves much faster to higher values for increasing salt concentrations. However, by shifting the storage modulus along the time axis, $G'(t)$ can be rescaled over a wide range of salt concentrations into a master curve with a remarkable sigmoidal shape in logarithmic scales. Such time-concentration superposition is robust to changes in the nature of the salt cation and in the CNC concentration. This points towards some universality in the recovery of the microstructure of CNC gels. 

Overall, the salt content sets the rate at which a robust microstructure is formed, whose long-term elastic properties are controlled by the CNC concentration and, to a lesser extent, by the nature of the counterion. Time-composition superposition further suggests that, for a given position along the universal master curve, i.e., for a given effective age, the interparticle interactions and the topology of the gel network should be very similar, independent of the CNC concentration or salt nature. We have confirmed this hypothesis by showing that both linear and nonlinear observables extracted from the viscoelastic spectra and from LAOS tests all follow the same behavior when plotted against the effective sample age $t_w/t^*$, where $t_w$ is the waiting time at rest and $t^*$ is the time at which $G'(t)$ reaches the inflexion point. The universality revealed in the present work opens up the following questions.

\subsection{Open questions}

\subsubsection{What is the functional form of the universal master curve?}

Master curves for the evolution of the storage modulus during gelation have been obtained in other colloidal systems with shapes similar to those of Figs.~\ref{fig:serie_NaCl}(b) and \ref{fig:comparaison_sels}(a). For instance, in some thermoreversible gels of sterically stabilized silica particles, it has been deduced from time-temperature superposition that the storage modulus at infinite time is independent of the temperature \cite{Negi.2014}. Here, we may deduce from time-composition superposition that the final storage modulus $G'(t=\infty)$ of CNC gels is independent of the salt concentration.

Moreover, master curves for $G'(t)$ are often fitted to exponential functional forms, $G'(t)=G'_\infty(1-\exp[-\lambda(t/t_g-1)^q])$, where $\lambda$  quantifies the rate of increase of connectivity of the gel, and $t_g$ is the ``gelation time,'' defined as the time when a critical gel is first observed \cite{Winter:1997} or sometimes more pragmatically as the time when the sample becomes solidlike, i.e., when $G'>G''$, for a given frequency \cite{Negi.2014} (see Fig.~\ref{fig:relax_gprime_gsecond_12mM_nacl}). Note that this expression implies that $G'(t_g)=0$, which is {\it a priori} not compatible with the definition of $t_g$, but still leads to realistic fits of the data, since the storage modulus at the gel point takes very small values. The exponent $q=1$, i.e., a simple exponential function, was reported to fit the evolution of $G'(t)$ in the above-mentioned thermoreversible silica gels \cite{Rueb:1997,Guo:2011,Negi.2014}, while salt-induced gels of Ludox silica particles yielded values of $q=1.6$--2.1, i.e., \emph{compressed} exponentials, depending on the colloid volume fraction \cite{Cao.2010}. In the case of CNC gels, we have shown that a \emph{stretched} exponential with $q\simeq 0.16$ accounts well for the $G'(\tilde{t})$ master curve, although with a slightly different form, $G'(\tilde{t})=G'_{\infty}(1-A \ \exp[-((\tilde{t}-\tilde{t}^*)/T_{q})^{q}])$  involving an additional fitting parameter $A$, and provided the characteristic time is taken as $\tilde{t}^*$ rather than the much smaller relative gelation time $t_g/\tau$ [see red line in Fig.~\ref{fig:serie_NaCl}(b)]. 

However, a single exponential form only captures the later stages of the aging dynamics for $\tilde{t}\gtrsim\tilde{t}^*$ and cannot reproduce the sigmoidal shape of the master curve as observed in logarithmic scales in Figs.~\ref{fig:serie_NaCl}(b) and \ref{fig:comparaison_sels}(a). Although we could fit the initial recovery and aging regime for $\tilde{t}\lesssim\tilde{t}^*$ by a stretched exponential growth with an exponent $p=0.28$ [see blue line in Fig.~\ref{fig:serie_NaCl}(b)], a single functional form fitting the whole dynamics is still lacking. In particular, the fact that $G'(t)$ does not tend to 0 at short times is indicative of some non-negligible initial elasticity immediately after preshear. This small level of elasticity could be due to the incomplete disaggregation of the CNC clusters by shear or to some intrinsic viscoelasticity of the fully dispersed CNC suspension that could result from partial liquid-crystalline order. More experiments are needed to clarify the origin of the initial elastic modulus upon flow cessation, e.g., by systematically varying the preshear value $\dot\gamma_p$. The physico-chemical parameters that control the amplitude and the steepness of the sigmoid in the master curve also remain to be uncovered.

Finally, the above discussion raises the question of the interpretation of the stretching exponents observed here, $p\simeq 0.28$ and $q\simeq 0.16$, which contrast with the compressed exponentials of Ref.~\cite{Cao.2010}. Stretched exponentials are typically reported from the relaxation of correlation functions measured in glassy systems where particles are enclosed in cages formed by their neighbors \cite{Cipelleti:2003}. On the other hand, systems where thermal motion dominates over internal constraints rather display governed by compressed exponentials \cite{Cipelletti:2002,Fluerasu:2007,Bouzid:2017}. Whether or not the stretched exponentials reported here in $G'(t)$ result from the glassy-like dynamics of clusters of CNCs remains to be investigated.

\subsubsection{How may one rationalize the influence of salt on the dynamics?}

In this work, the ``ìnflexion time'' $t^*$ provides a characteristic time for the formation of the gel microstructure, which we may compare to the ``gelation time'' that is more classically reported in the literature. We have shown that, for an ionic strength below 100~mM, $t^*$ decays as a power law of the ionic strength $I$ with an exponent $\beta=8.7\pm 0.2$ independent of the nature of the cation and of the CNC concentration. Interestingly, a similar, very steep exponent of about $-10$ has been found in gels of cotton CNC for the dependence upon salt concentration of the gelation time $t_g$ inferred from the intensity of light scattering \cite{Peddireddy:2016}. Moreover, visual estimations of the sol-gel transition time in Ludox silica gels have reported power laws $t_g\sim I^{-\beta}$ with $\beta=5.9$--8.4 depending on the type of salt and on the colloid volume fraction \cite{Linden.2015}. Power-law behaviors with exponents in the range $\beta=6$--11 for the stability ratio $W$, or equivalently for the inverse of the coagulation rate, as a function of salt concentration have been predicted theoretically for monodisperse spherical particles and indeed found in pioneering experiments on AgI colloids through turbidity measurements \cite{Reerink.1954}. Since the gelation time is directly proportional to the stability ratio \cite{Zaccone.2014,Linden.2015}, similar exponents are expected for $t_g$ vs $I$. The theory however predicts that the exponent should depend on the salt, in particular on the valency $z$ of the counterion \cite{Reerink.1954,Linden.2015}, whereas we do not observe any significant variation in $\beta$ with the type of salt. This discrepancy could be ascribed to the non-spherical nature and/or to the polydispersity of CNCs. To the best of our knowledge, a complete theory that would account for the specific characteristics of CNCs, both in terms of geometry and of surface charges, and for their interactions in the presence of salt is yet to be devised.
 
While $\beta$ seems to remain constant, we have reported in Sect.~\ref{sec:comparaison_sels} a significant dependence of the prefactor $B$ in $t^*=B/I^{\beta}$, with the type of counterion. In particular, for a fixed CNC content, $B$ increases when going from \ch{K+} to \ch{Na+} and then to \ch{Mg^{2+}}. Thus, the kinetics become slower as one follows the Hofmeister series from more chaotropic (or ``structure-breaking'') ions, such as \ch{K+}, to more kosmotropic (or ``structure-forming'') ions, such as \ch{Mg^{2+}}, which surround themselves with a greater number of water molecules \cite{Kunz:2004a,Kunz:2004b}. To explain such a dependence, one must complement the classical DLVO potential with an additional ion-specific repulsive potential \cite{Israelachvili:1996} due to the hydration shell that surrounds ions adsorbed on the CNC surfaces, as proposed for silica particles not only for the gelation times \cite{Linden.2015} but also for the rheological properties of the resulting gels \cite{Franks.2002,Okazaki.2008}. Kosmotropic ions have a greater hydration diameter, which makes their adsorption on the colloid surface more difficult,
leading to a larger effective hydration repulsion. This scenario qualitatively explains the slowing down of the gelation and aging kinetics for more kosmotropic ions, although a full quantitative interpretation for highly charged rodlike particles such as CNCs still remains out of reach.

\subsubsection{How does the microstructure evolve under aging?}

Besides the influence of salt, the volume fraction of CNCs plays a crucial role in the recovery and aging dynamics, since an increase of 50~\% in the CNC content from 3.2~wt.~\% to 4.8.~\% accelerates the kinetics by a factor of about 100 (see Sect.~\ref{sec:comparaison_sels}).
Such an acceleration when increasing the CNC volume fraction has also been reported from turbidity measurements, although these were limited to the recovery phase \cite{Peddireddy:2016}. This suggests that the more particles, the more interactions, and the sooner the equilibrium configuration is reached. This also questions the role of interactions at the molecular scale between CNC clusters, including hydrogen bonds as recently emphasized in Refs.~ \cite{Rao.2019,Wohlert:2021}.

From the mechanical measurements presented in Sect.~\ref{sec:viscoelastic}, we may further elaborate on the potential gel structure reached along the $G'(t)$ master curve. In particular, in the framework of critical gels, originally developed for branched polymer gels \cite{Muthukumar:1989,Winter:1997,Ng:2008} and later extended to silica polymers \cite{Ponton:2002}, protein gels \cite{Ikeda:2001,Ikeda:2003}, fibrin-thrombin gels \cite{Curtis:2013}, thermoreversible gels of silica nanoparticles \cite{Eberle:2012}, or alumino-silicate gels \cite{Keshavarz.2021}, the exponent $\alpha$ that characterizes the frequency-dependence of the viscoelastic spectrum has been linked to the fractal dimension $d_f$ of the particulate network through $\alpha=3(5-2d_f)/2(5-d_f)$, under the assumption that hydrodynamic and excluded-volume interactions are fully screened. In our case, this relationship leads to a fractal dimension that increases from $d_f=1.4$ to about 2.5 with the salt content, or, equivalently, with the effective sample age. These estimates are fully consistent with previous results on similar gels of charged cotton CNC rods in the presence of salt, which report fractal dimensions $d_f\simeq 1.6$ for a moderate salt concentration (70~mM NaCl) using light scattering \cite{Peddireddy:2016} and $d_f\simeq 2.1$ at high ionic strength (200~mM NaCl) through small-angle neutron scattering \cite{Cherhal:2015}.

The nonlinear viscoelastic measurements of Sect.~\ref{sec:nonlinear} provide additional insight into the structural evolution of CNC gels under aging. When moving along the master curve, i.e., when equivalently considering longer waiting times or larger salt concentrations, both characteristic strains $\gamma_{\rm NL}$ and $\gamma_{\rm c}$ decrease dramatically, suggesting that the gel becomes more and more ``brittle'' when aging. A similar drop in $\gamma_{\rm NL}$ and $\gamma_{\rm c}$ has been reported in cellulose gels for increasing salt concentration \cite{Moud.2020}. It was attributed to the formation of a much stronger network due to denser clusters, which is consistent with the possible increase of the fractal dimension $d_f$ discussed above. Still, structural measurements along the master curve are required to confirm the fractal nature of the present CNC gels and the evolution of $d_f$ with the effective sample age. Ideally, time-resolved structural measurements coupled to LAOS tests will provide a full picture of the microstructure of CNC gels during yielding as a function of sample age.


\subsubsection{How do dissipative processes scale during the aging dynamics?}

Although we mostly focused our analyses on the storage modulus $G'$, the loss modulus $G''$ and the loss tangent $\tan\delta$ also carry important information on the aging dynamics, and more particularly on the way dissipation occurs throughout the aging process. One direction for a deeper investigation of dissipative processes concerns the evolution of the $G''$ overshoot during yielding. The specific ``type III'' fluidization scenario reported in Fig.~\ref{fig:SS}(a) is reminiscent of the yielding transition reported in soft glasses made of hard-sphere colloids or jammed emulsions \cite{Mason:1995c,Rogers:2018,Pham:2006}, or weak polymer gels involving stiff or charged molecules \cite{Hyun.2002}. The overshoot in $G''$, known as the Payne effect in the context of rubber, has been associated with the increased dissipation due to irreversible, plastic deformation \cite{Mason:1996a,Rogers:2018} and recently quantified through time-resolved decomposition of recoverable and unrecoverable strains \cite{Donley.2020}.

Here, we have shown that the relative amplitude of the $G''$ overshoot, $\Delta G''/G''_0$, depends on the sample age (or on the salt concentration) in a non-monotonic fashion. Although a decrease of $\Delta G''/G''_0$ for increasing salt concentration has been reported in a dense assembly of microgels \cite{Shao.2013}, this is, to our knowledge, the first time that a non-monotonic dependence is reported. In particular, the salt concentrations at which we observe the largest overshoots in $G''$ coincide with the compositions for which $G'(t)$ goes through the inflexion point of the master curve [see Fig.~\ref{fig:serie_NaCl}(a)]. This suggests that the fast aging of the sample contributes to enhance the viscous dissipation at the yield point.

Finally, note that a microscopic interpretation of the $G''$ overshoot in CNC gels has been proposed recently based on the intracycle analysis of LAOS measurements \cite{Moud.2020}: the increase in strain amplitude would reduce the interparticle distance, thus allowing the formation of ``shear-induced networks'' that increase the viscous dissipation when dragged along by shear. There again, time-resolved structural measurements under LAOS would be necessary to confirm such an interpretation.


\subsection{Concluding remarks and perspectives}

The results presented in this study clearly highlight some universality in the recovery and aging dynamics of CNC gels. As illustrated by the above open questions, our mechanical measurements pave the way for future investigations, including a time-resolved characterization of the microstructure, e.g. through imaging under polarised light or through small-angle light scattering, in order to identify the microscopic mechanisms at play in such slow dynamics. More generally, we expect that experimental results combining mechanical and microstructural characterization will feed numerical simulations and theoretical modelling, in order to fully understand the physics of time-composition superposition in charged rodlike colloids.

\subsection*{Acknowledgements}
The authors are grateful to E.~Freyssingeas, T.~Gibaud, B.~Jean, F.~Pignon, and J.-L. Puteaux for fruitful discussions on the physico-chemistry of CNCs and on the gel microstructure.

%


\clearpage
\newpage
\onecolumngrid
\setcounter{equation}{0}
\setcounter{figure}{0}
\global\def\thefigure{S\arabic{figure}}
\setcounter{table}{0}
\global\def\thetable{S\arabic{table}}

\onecolumngrid

\begin{center}
    {\large\bf Slow dynamics and time-composition superposition in gels of cellulose nanocrystals}
\end{center}

\begin{center}
    {\large\bf{\sc Supplemental Material}}
\end{center}

\section*{Temporal evolution of the loss modulus after shear rejuvenation}
\label{app:loss_modulus}
\begin{figure*}[h!]
    \centering 
    \includegraphics[width=0.9\textwidth]{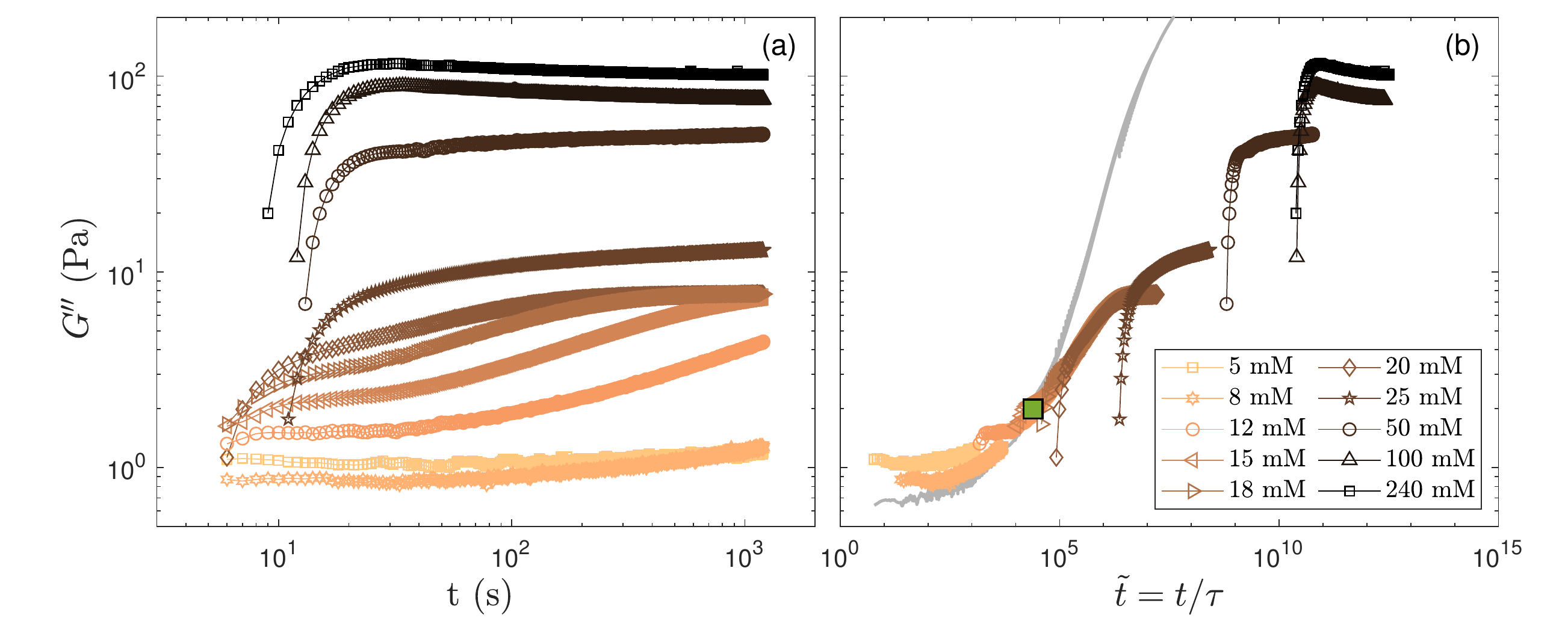}
    \caption{\label{fig:annexes_Gsecond_NaCl} (a) Temporal evolution of the loss modulus $G''$ following a $20$~s preshear at $\gp=500$~s$^{-1}$. Experiments performed on 3.2~wt.~\% CNC gels with NaCl concentrations ranging from 5~mM to 240~mM (colored symbols). (b)~Master curve obtained by shifting the $G''(t)$ data in (a) along the time axis by a factor $1/\tau$. The response $G''(t)$ of the gel with 5~mM of NaCl is used as a reference ($\tau=1$~s). The gray curve shows the $G'(\tilde t)$ master curve pictured  in Fig.~2(b) in the main text, and the green square highlights the point at which $G'$ and $G''$ cross.}
\end{figure*}

Figure~\ref{fig:annexes_Gsecond_NaCl}(a) displays the temporal evolution of the loss modulus $G''(t)$ corresponding to the storage modulus $G'$ shown in Fig.~2(a) in the main text.
Similarly to $G'$, the loss modulus $G''$ increases with time following preshear cessation, and the kinetics strongly depend on the salt concentration. 
Using the same shift factors $\tau$ as for $G'$, the data $G''(t)$ for different salt concentrations leads to the master curve of Fig.~\ref{fig:annexes_Gsecond_NaCl}(b). The rescaling works rather well for low salt concentrations, over about three orders of magnitude around the crossing point of $G'(\tilde t)$ and $G''(\tilde t)$ [marked by a green square in  Fig.~\ref{fig:annexes_Gsecond_NaCl}(b)]. However, for samples with a salt concentration larger than 18~mM, the short-time response of $G''(t)$ following flow cessation shows a clear departure from the hypothetical master curve. Such a discrepancy could be linked to the specific way flow cessation is triggered, or to an abrupt change in the dissipation process upon flow cessation. 
Moreover, we note that for the two largest salt concentrations, $G''(t)$ shows an overshoot before decreasing at long times. This non-monotonic response, which is absent from $G'(t)$, could be related to the syneresis of the gels, i.e., the spontaneous expulsion of solvent that is observed when these two specific samples are left at rest. The issue of syneresis in CNC gels certainly deserves more attention in view of its potential importance for applications.  

\vspace{2cm}

\clearpage


\section*{Master curves for different CNC contents and salt concentrations}
\label{app:storage_modulus}

\begin{figure*}[!h]
    \centering 
    \includegraphics[width=0.9\textwidth]{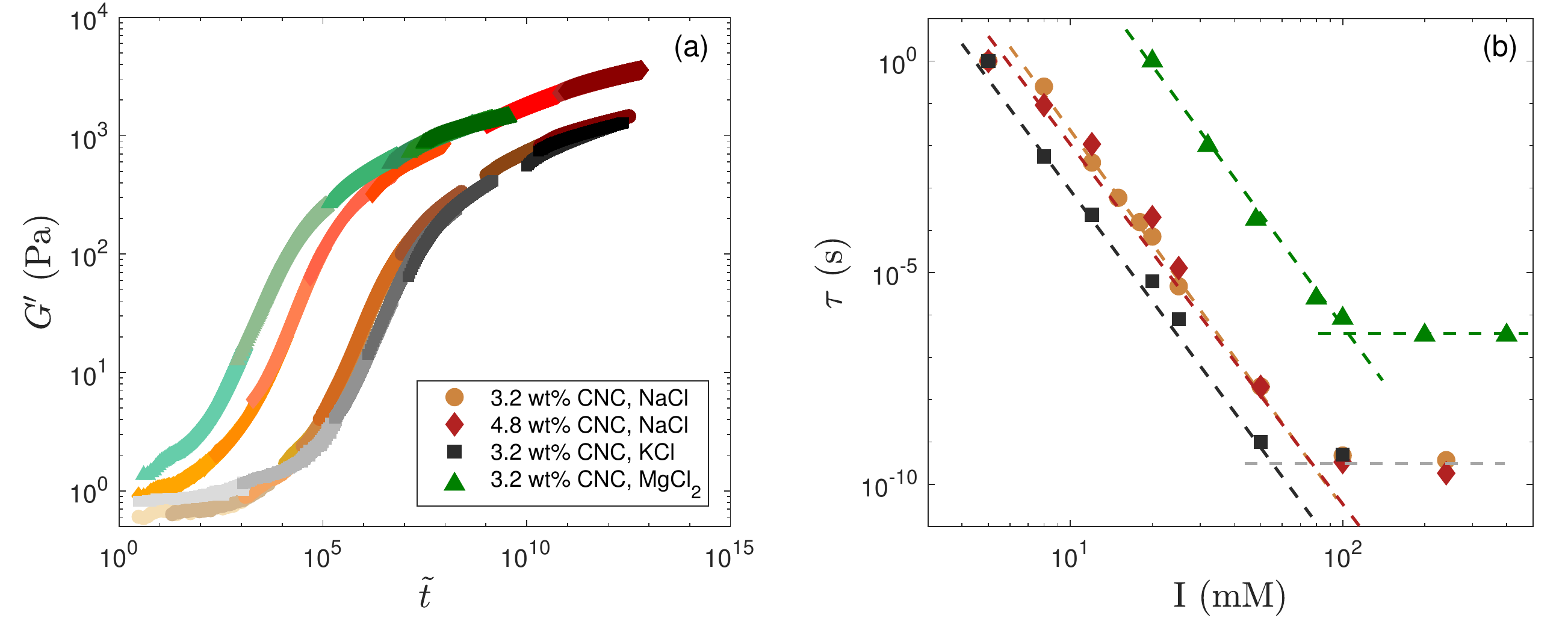}
    \caption{\label{fig:annexes_Gprime_tau} (a)~Storage modulus $G'$ vs.~$\tilde t=t/\tau$ during recovery and aging following a $20$~s preshear at $\gp=500$~s$^{-1}$. The response $G'(t)$ of the gels with 5~mM of salt are used as references ($\tau=1$~s). Experiments performed on four different series of samples containing either 3.2 or 4.8\% of CNC, and different types of salt, namely NaCl, KCl or \ch{MgCl2}, at concentrations ranging from 5 to 240~mM. (b)~Shift factors $\tau$ vs.~the ionic strength $I$ for the different series of gels [same symbols as in (a)]. Dashed lines correspond to the power-law fits of $\tau$ vs. $I$ for $I<100$~mM. The values of the exponents are reported in Table~1 in the main text. Horizontal dashed lines highlight the plateaus reached beyond $I\simeq 100$~mM.}
\end{figure*}

We apply the same rescaling scheme as in Fig.~2 in the main text to the $G'(t)$ data associated with four series of samples containing different CNC concentrations and different types of salt, namely NaCl, KCl and \ch{MgCl2} at various concentrations. 
Figure~\ref{fig:annexes_Gprime_tau}(a) shows that for all CNC concentrations and types of salt, $G'(t)$ can be rescaled onto a master curve similar to the one presented in Fig.~2(b) in the main text. The shift factor $\tau$ reported in Fig.~\ref{fig:annexes_Gprime_tau}(b) evolves similarly with the ionic strength $I$ for the four series of gels: it first decreases following a power law, until it reaches a plateau for ionic strengths above 100~mM. 

\begin{figure}[b]
    \centering 
    \includegraphics[width=0.45\textwidth]{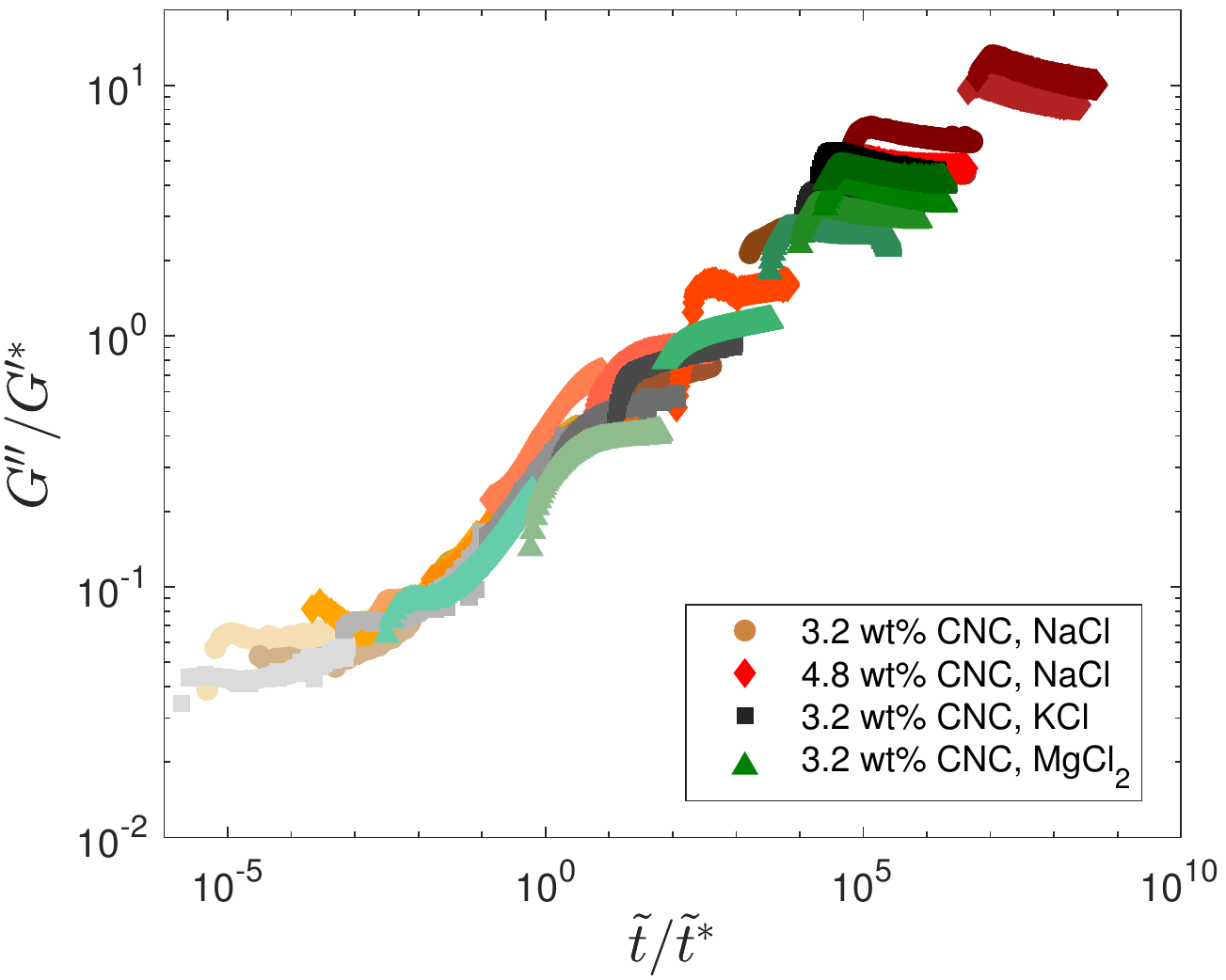}
    \caption{Normalized loss modulus $G''/G'^*$ vs.~normalized time $\tilde t/\tilde t^*$ during recovery and aging following a $20$~s preshear at $\gp=500$~s$^{-1}$. Experiments performed on four different series of samples containing either 3.2 or 4.8\% of CNC, and different types of salt, namely NaCl, KCl or \ch{MgCl2}, at concentrations ranging from 5~mM to 240~mM. The values of $G'^*$ and $\tilde t^*$ for each series of samples are reported in Table~I in the main text.} \label{fig:annexes_Gsecond_tous_sels}
\end{figure}

The master curve obtained when plotting the loss factor $G''$ rescaled by $G'^*$ as a function of $\tilde{t}/\tilde{t}^*$ is shown in Fig.~\ref{fig:annexes_Gsecond_tous_sels}. As already mentioned above, the rescaling of $G''$ does not hold as well as for $G'$, especially at large effective ages, i.e., at large salt concentrations.

\vspace{0.5cm}

\clearpage

\section*{Analysis of strain sweeps at different salt concentrations}
\label{app:strain_sweep}

\begin{figure*}[!h]
    \centering 
    \includegraphics[width=0.9\textwidth]{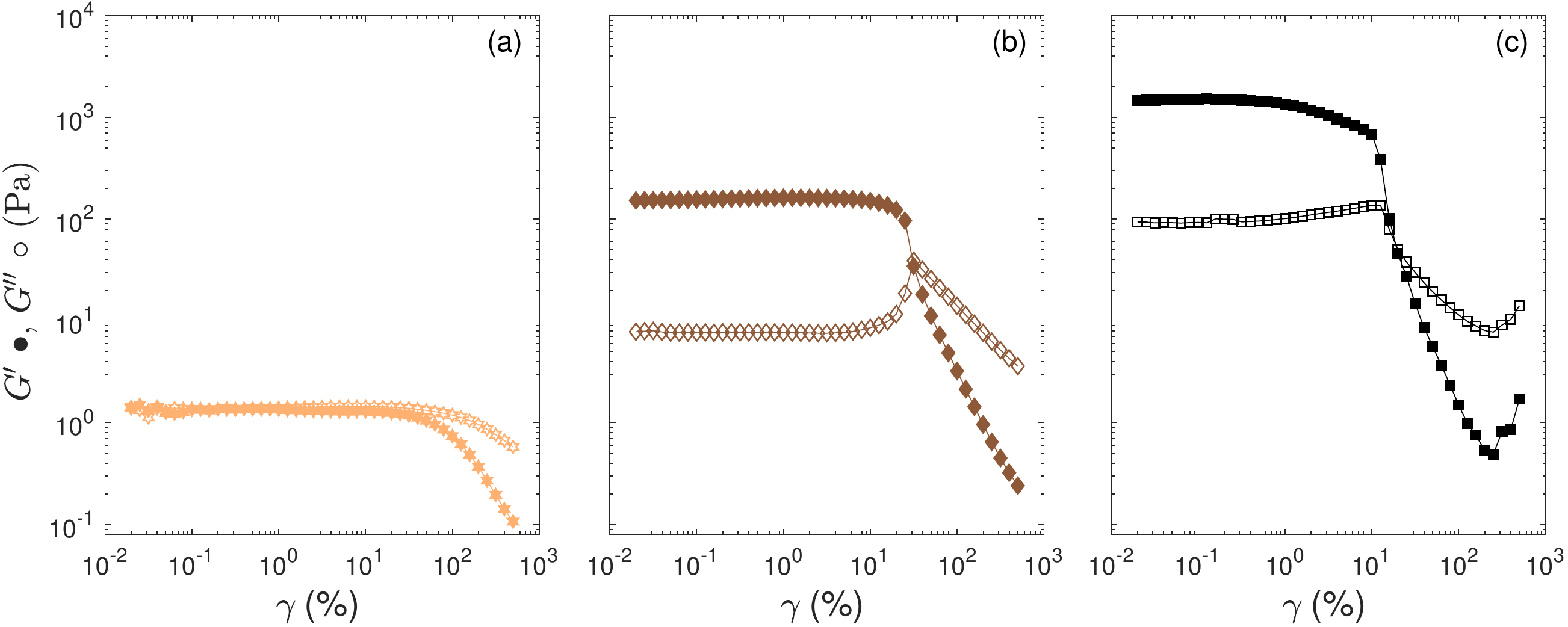}
    \caption{\label{fig:annexes_SS} Storage modulus $G'$ and loss modulus $G''$ vs. strain amplitude $\gamma$ following $1470$~s of rest after a $20$~s preshear at $\gp=500$~s$^{-1}$. Experiments performed on samples containing 3.2~wt.~\% of CNC and NaCl at (a) 8~mM, (b) 20~mM, and (c) 240~mM.} 
\end{figure*}

Figure~\ref{fig:annexes_SS} shows oscillatory strain sweeps performed on 3.2 wt.~\% CNC gels with three different salt concentrations, 8, 20 and 240~mM, spanning the whole range of salt content under study.
While the gel containing 8~mM NaCl is barely solidlike at rest ($G'\simeq G''$) and displays a continuous transition from the linear to the nonlinear regime, 
the gels containing 20~mM and 240~mM NaCl exhibit an abrupt yielding transition characterized by an overshoot in $G''$, and a power-law dependence of both $G'$ and $G''$ with the strain beyond the crossing point. 

\begin{figure}[h]
    \centering 
    \includegraphics[width=0.5\textwidth]{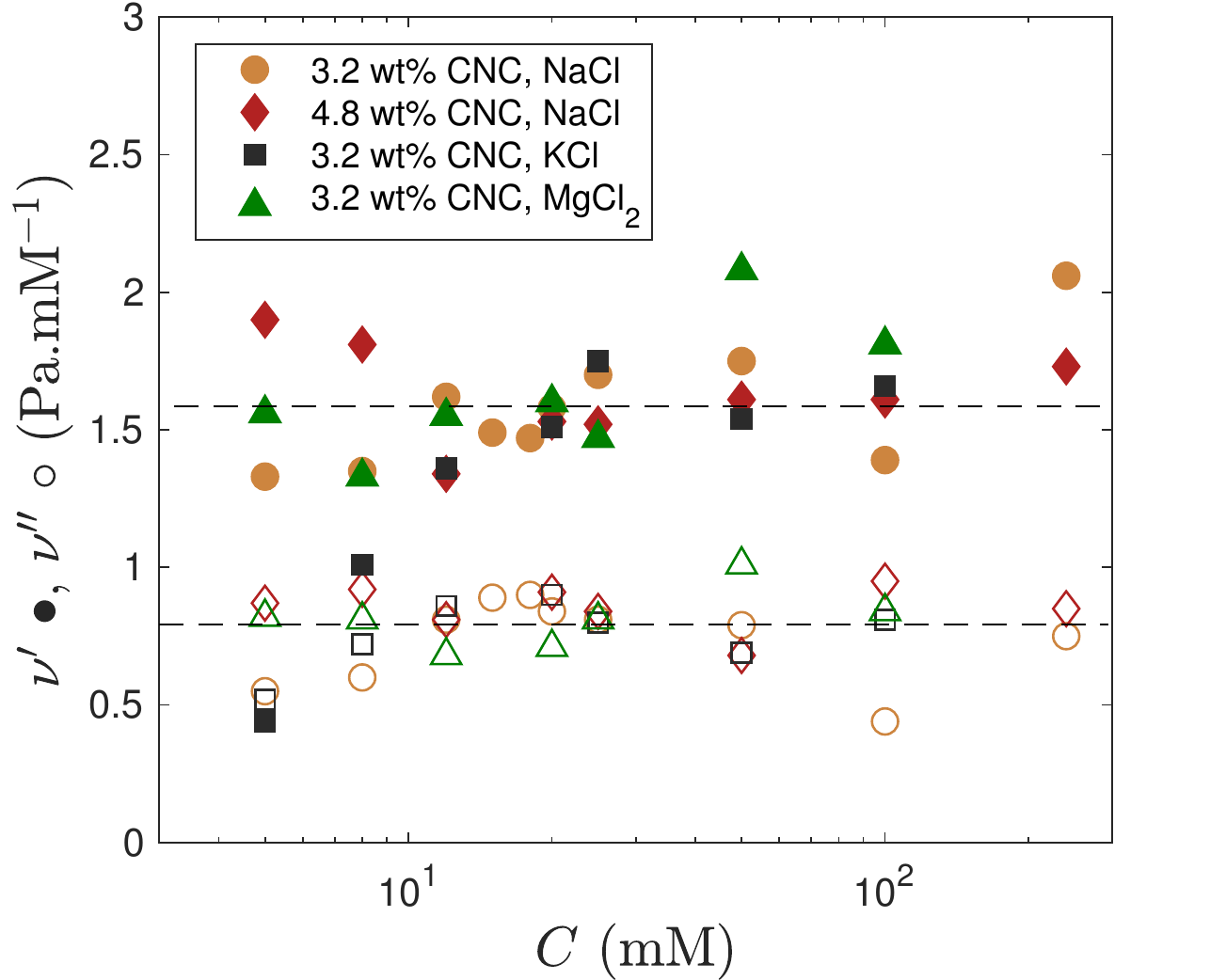}
    \caption{Exponents $\nu'$ and $\nu''$ of the power-law decrease of $G'$ and $G''$ respectively, observed beyond the yield point during oscillatory strain sweeps vs.~salt concentration $C$. Experiments performed on samples containing either 3.2 or 4.8\% of CNC, and different types of salt, namely NaCl, KCl or \ch{MgCl2}, at concentrations ranging from 5 to 240~mM.} \label{fig:annexes_pentes_SS} 
\end{figure}

Figure~\ref{fig:annexes_pentes_SS} reports the exponents $\nu'$ and $\nu''$ associated with the power-law decrease of $G'$ and $G''$ with the strain amplitude beyond the yield point as a function of salt concentration [see Fig.~5(a) in the main text for the definition of $\nu'$ and $\nu''$]. The data gather four different series of samples containing various CNC concentrations or types of salt as in Figs.~3 and 4 in the main text. These two exponents depend neither on the concentration or type of salt, nor on the CNC concentration: $\nu' \simeq 1.6$ and $\nu'' \simeq 0.8 \simeq \nu'/2$.

\begin{figure}[h]
    \centering 
    \includegraphics[width=0.5\textwidth]{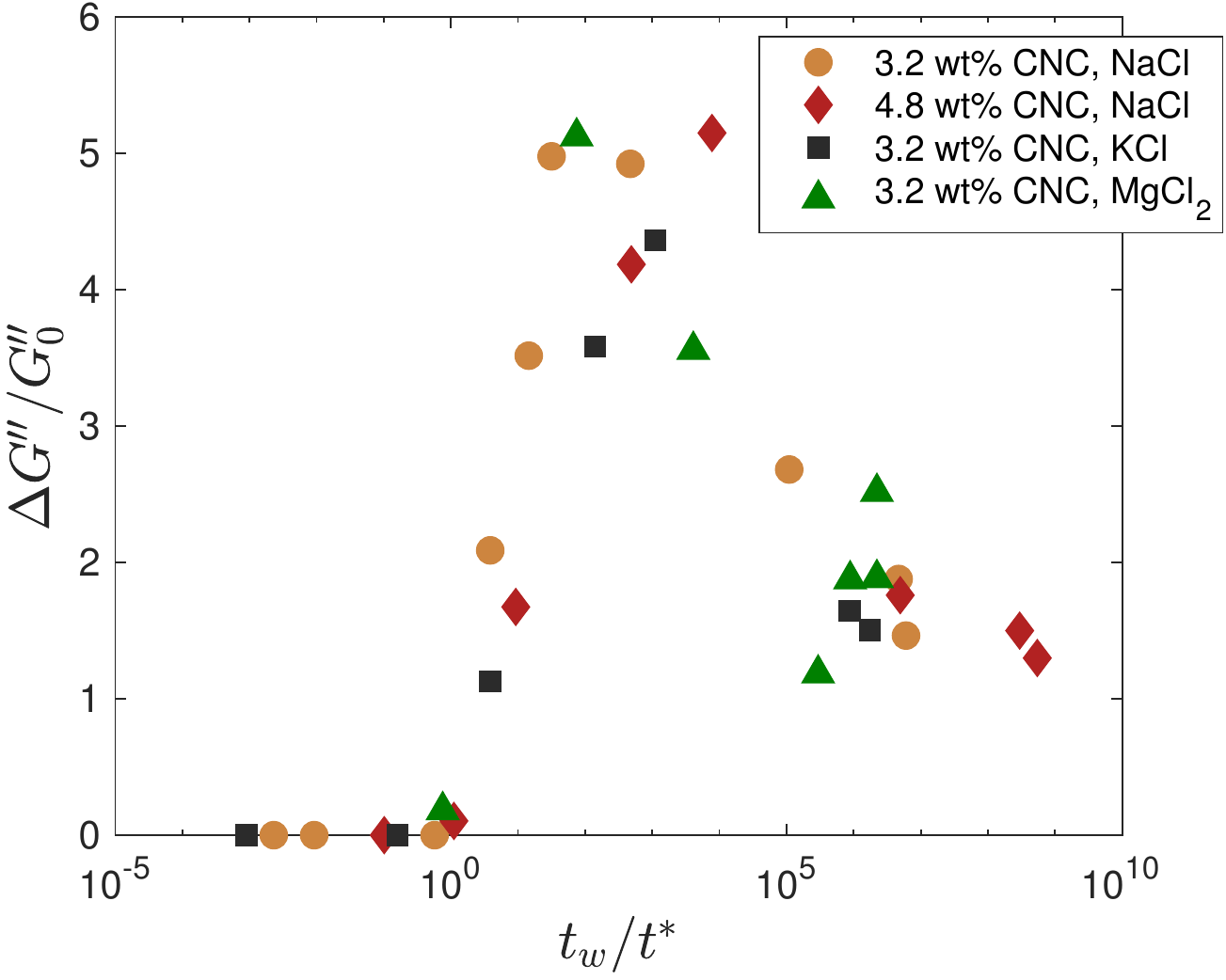}
    \caption{Dimensionless amplitude of the loss modulus overshoot $\Delta G''/G''_0$ as a function of the effective age $t_w/t^*$. $G''_0$ is the plateau value of the loss modulus measured in the linear regime, i.e., at small strain amplitude. Experiments performed on samples containing either 3.2 or 4.8\% of CNC, and different types of salt, namely NaCl, KCl or \ch{MgCl2}, at concentrations ranging from 5 to 240~mM.} \label{fig:annexes_overshoot_Gsecond_SS}
\end{figure}

Figure~\ref{fig:annexes_overshoot_Gsecond_SS} shows the dimensionless amplitude of the loss modulus overshoot $\Delta G''/G''_0$ observed at the yield point during an oscillatory strain sweep as a function of the effective age $t_w/t^*$ for samples containing either 3.2 or 4.8\% of CNC, and different types of salt, namely NaCl, KCl or \ch{MgCl2}, at various concentrations. The non-monotonic behavior of $\Delta G''/G''_0$ with $t_w/t^*$ is an outstanding feature of the $G''$ overshoot in CNC gels.

\vspace{2cm}

\end{document}